# On two classes of perturbations: Role of signs of second-order Rayleigh-Schrödinger energy corrections

Kamal Bhattacharyya*
Department of Chemistry, University of Calcutta, Kolkata 700009, India

**Abstract**

We distinguish two extreme classes of perturbation problems depending on the signs of second-order energy corrections and argue why it is generally much more probable to obtain a negative value of the same for any state in the standard Rayleigh-Schrödinger perturbation theory. The classes are seen to differ in reproducing results of finite-dimensional matrix perturbations. A few related issues are also discussed, some of which are based on available analytical results.

*e-mail: pchemkb@yahoo.com; pchemkb@gmail.com

## 1. Introduction

It is well known in non-degenerate Rayleigh-Schrödinger perturbation theory (RSPT) [1 – 3] that the second-order energy correction term is negative for the *ground state only*. In general, for some *n*-th state, this correction term takes the form

$$E_{n2} = \sum_{m \neq n}^{\infty} \frac{|V_{mn}|^2}{E_{n0} - E_{m0}} \tag{1}$$

where $V_{mn} = \langle \Psi_{m0} | V | \Psi_{n0} \rangle$, $H_0 \Psi_{m0} = E_{m0} \Psi_{m0}$ and $H = H_0 + \lambda V$. Here, $H$ is the full system Hamiltonian and $H_0$ is the unperturbed one. The eigenenergy $E_n$ of $H$ is given by a power series in $\lambda$:

$$E_n = E_{n0} + \lambda E_{n1} + \lambda^2 E_{n2} + \ldots \tag{2}$$

It is customary to distinguish perturbation problems on the basis of whether (2) is convergent or divergent. But, here we shall take a different route based on the signs of $E_{n2}$. It does not follow from (1) that $E_{n2}$ ($n > 0$) will have any *specific* sign that is irrespective of the *nature* of the problem. So, one may look for generalities like when *all* $E_{n2}$ ($n > 0$) will have a *definite* sign, either positive or negative, and, if so, how probable are these two extreme varieties. It is important to gather some such information because this quantity is directly related to the *primary response property* of the unperturbed system for a given perturbation in the concerned quantum state. Polarizability and susceptibility are two most significant properties of this kind. Additionally, one can see how these two classes of problems differ in approving finite-dimensional matrix perturbation results. This is another interesting feature, because normally one does not expect infinite-dimensional problems to show up characteristics of finite-dimensional ones.

Employing (1), one obtains the interesting inequality [3]

$$\sum_{n=0}^{K} E_{n2} < 0, \tag{3}$$

for any $K \geq 0$. This is easy to verify, but (3) does not seem to have attracted much attention. In case of finite-dimensional [$(N + 1)$] matrix perturbation problems, one may note that the inequality in (3) would hold for any $K < N$; only for $K = N$ we have the equality:

$$\sum_{n=0}^{N} E_{n2} = 0. \tag{4}$$

Therefore, it may be taken for granted generally that, for quantum-mechanical problems, where $N \to \infty$, we would have the *inequality* in (3) valid for any K. But, we shall notice that a class of Hilbert space problems does satisfy (4), contrary to common sense [4].

Specifically, our investigations on the above two questions reveal that (i) it is much more probable to obtain a negative second-order energy correction for any state in RSPT and (ii) if $E_{n2}$ ($n > 0$) turns out to be *all positive, even excluding* the first few low-lying states, the equality in (4) is satisfied.

A few other features are worth mentioning. First, the signs of $\{E_{n2}\}$ cannot vary arbitrarily because they have to satisfy (3). However, the chance of finding a positive $E_{n2}$ apparently increases with increasing $n$ due to the involvement of a larger number of positive terms in the sum (1). Second, cases where $E_{n2}$ can be obtained in closed forms provide some insight into potential families that always yield a second-order shift of a definite sign for any state. Indeed, here lies the interest behind studies on simple systems that offer exact analytic results for $E_{n2}$. Third, we find it coincidentally that the agreement between finite [$(N + 1)$] and infinite [$N \to \infty$] dimensional results regarding the equality in (4) occurs only when we have a set of *all positive* $E_{n2}$, at least for large $n$. Such an outcome can be proved and, in situations, are verifiable.

Motivation behind a study of this sort emerges out of a curious observation on a large number of model Hamiltonians for which exact $E_{n2}$ are calculable. For example, perturbative studies [5 – 7] on the harmonic oscillator show that the Hamiltonian family

$$H = -\frac{1}{2}\frac{d^2}{dx^2} + x^2 + \lambda x^M. \tag{5}$$

yields the feature $E_{n2} < 0$ for *any M* and at any *n*. Here are a few results of $E_{n2}$ *at large n* where one normally expects otherwise:

$$M = 3: -\frac{7}{4}n^2; \quad M = 4: -\frac{17}{4}n^3;$$
$$M = 5: -\frac{187}{16}n^4; \quad M = 6: -\frac{393}{16}n^5. \tag{6}$$

The trend is clear; with increasing M, $E_{n2}$ tends to be more negative. RSPT studies [7, 8]

on the hydrogenic s-state Hamiltonian of the form

$$H = -\frac{1}{2}\frac{d^2}{dr^2} - \frac{1}{r}\frac{d}{dr} - \frac{1}{r} + \lambda r^M \tag{7}$$

leads to similar trends of $E_{n2}$ for *any M*. A few representative results in the large-*n* limit are as follows:

$$M = 1: -\frac{7}{8}n^6; \quad M = 2: -\frac{143}{16}n^{10};$$
$$M = 3: -\frac{7365}{128}n^{14}; \quad M = 4: -\frac{80123}{256}n^{18}. \tag{8}$$

Once more, we find here that $E_{n2} < 0$ for *any n*. It is also seen from (8) that $E_{n2}$ will only be more negative for larger *M*. Since the harmonic oscillator and H atom problems are two primary exactly-solvable problems in quantum mechanics, and perturbations of the form (5) or (7) constitute an infinitude of possibilities, one is inclined to believe that some kind of generality should involve the observation $E_{n2} < 0$ ($n > 0$). The only exception, to the best of our knowledge, is the particle-in-a-box (PB) perturbed by a specific (linear) perturbation [9, 10]. It is, therefore, imperative to provide a general proof of the *prominence* of $E_{n2} < 0$, at least under certain constraints.

Since we choose a very general, statistical approach, our main results have little to do with analytically solvable problems; however, a few other findings do have their roots in exactly obtainable corrections.

**2. The main results**

Denoting $E_{02}$ by $-\alpha$ ($\alpha > 0$), we obtain from (3) for $K = 1$

$$E_{12} < \alpha. \tag{9}$$

More specifically, (9) can be written as

$$E_{12} = \beta_1\alpha, \, \beta_1 < 1. \tag{10}$$

For $K = 2$ in (3), one can likewise write

$$E_{22} = \beta_2(1-\beta_1)\alpha, \, \beta_2 < 1. \tag{11}$$

In general, therefore, it follows that

$$E_{k2} = \beta_k(1-\beta_{k-1})...(1-\beta_1)\alpha, \, \beta_k < 1, \tag{12}$$

where the preceding equations define the earlier $\beta_j$. Form (12) will be quite useful for our purpose.

We now cast our problem in the following way. Suppose *all* $E_{k2}$ terms ($k > 0$) are positive or zero. This implies, $\beta_k$ is bounded as $0 \leq \beta_k < 1$. In such a situation, one can also order $E_{n2}$ as

$$|E_{02}| > E_{12} > E_{22} > E_{32}... \quad . \tag{13}$$

On the other hand, if *all* $E_{k2}$ terms ($k > 0$) are negative semidefinite, $\beta_k$ will be bounded as $-\infty < \beta_k \leq 0$. The domains (respectively finite and infinite) immediately point to the *overwhelmingly* larger probability of encountering the latter situation than the former, for arbitrary perturbations on arbitrary states.

One may next appreciate why the equality (4) will be valid even for infinite-dimensional cases, provided, of course, that *all* $E_{k2}$ terms for excited states are *positive semidefinite* (this point may further be relaxed, see below). Indeed, we can choose here, without any loss of generality,

$$\beta_k = \sin^2 \theta_k, \; 0 \leq \theta_k < \pi/2. \tag{14}$$

Then, it is easy to show that the left side of (4) takes the form

$$\sum_{k=0}^{K} E_{k2} = -\alpha \prod_{k=1}^{K} \cos^2 \theta_k. \tag{15}$$

But, for arbitrary $\{\theta_k\}$, the right side of (15) tends to zero as $K \to \infty$. Thus, for such a class of infinite-dimensional problems, it is *natural* that the *equality* in (4) will be obeyed.

The other possibility, on the contrary, is wide open. For example, if we choose $\{E_{k2}\}$ as *negative semidefinite* and similarly bounded (*i.e.*, $-\infty < \beta_k \leq 0$), it is permissible to take, instead of (14),

$$\beta_k = -\tan^2 \theta_k, \; 0 \leq \theta_k < \pi/2. \tag{16}$$

As a result, we are led to a sum of all negative terms

$$\sum_{k=0}^{\infty} E_{k2} = -\alpha \prod_{k=1}^{\infty} \sec^2 \theta_k, \; 0 \leq \theta_k < \pi/2. \tag{17}$$

One sees that, while a product term similar to (15) appears at (17), it now increases without limit for arbitrary $\{\theta_k\}$, *i.e.*, arbitrary perturbations, and hence the *inequality* in (3) will *always* hold. All such results are general ones and do not depend on the dimensionality (or, particle number) of the problem.

**3. Further remarks**

A few examples of RSPT with *all positive* $\{E_{k2}\}$ for excited states would now be chosen for demonstrative purposes. Consider the PB problem ($m = \frac{1}{2}$, $\hbar = 1$) in (0, 1) with a linear ($x$) perturbation. One obtains [9, 10], with $k$ referring to the quantum number that starts from unity,

$$E_{k2} = \frac{k^2\pi^2 - 15}{48k^4\pi^4}. \tag{18}$$

Hence, the relevant sum in (4) becomes

$$\sum_{k=1}^{\infty} E_{k2} = \frac{1}{48\pi^2}\left(1 + \frac{1}{2^2} + \frac{1}{3^2} + \ldots\right) - \frac{15}{48\pi^4}\left(1 + \frac{1}{2^4} + \frac{1}{3^4} + \ldots\right)$$

$$= \frac{1}{48\pi^2}\frac{\pi^2}{6} - \frac{15}{48\pi^4}\frac{\pi^4}{90} = 0. \tag{19}$$

A second example that we have constructed refers again to the above PB problem in the same domain, but with the perturbation $\cos(\pi x)$. It yields

$$E_{12} = -\frac{1}{12\pi^2}, \quad E_{k2} = \frac{1}{4\pi^2}\left(\frac{1}{2k-1} - \frac{1}{2k+1}\right), k > 1. \tag{20}$$

The required sum for the left side of (4) gives

$$\sum_{k=1}^{\infty} E_{k2} = \frac{1}{4\pi^2}\left[-\frac{1}{3} + \left(\frac{1}{3} - \frac{1}{5}\right) + \left(\frac{1}{5} - \frac{1}{7}\right) + \ldots\right]$$

$$= 0. \tag{21}$$

The above two results leading to (19) and (21) are concrete examples of (15). We now choose another case to justify (15) under more *relaxed* condition. Consider again the above PB problem with the perturbation $\cos(2\pi x)$. We have found in this case

$$E_{12} = \frac{1}{4\pi^2(1^2 - 3^2)}, \quad E_{22} = \frac{1}{4\pi^2(2^2 - 4^2)},$$

$$E_{k2} = \frac{1}{4\pi^2}\left(\frac{1}{k^2 - (k-2)^2} + \frac{1}{k^2 - (k+2)^2}\right), k > 2. \tag{22}$$

It follows immediately from (22) that

$$\sum_{k=1}^{\infty} E_{k2} = 0. \tag{23}$$

In fact, even if we have the *first few* $E_{k2}$ negative, but subsequent ones *all positive*, a proof like (15) can be easily found. For example, when the first three terms are negative, we obtain a result by combining (14) and (16) in the form

$$\sum_{k=0}^{\infty} E_{k2} = -\alpha \sec^2 \theta_1 \sec^2 \theta_2 + \alpha \sec^2 \theta_1 \sec^2 \theta_2 \left(1 - \prod_{k=3}^{\infty} \cos^2 \theta_k \right), \quad 0 \leq \theta_k < \pi/2 \qquad (24)$$
$$= 0.$$

Therefore, one can get *a class of problems* for which results of finite and infinite dimensions agree under conditions less restrictive than those involved in (15).

It is rather simple to see, *a posteriori*, why most one–dimensional perturbation problems that are analytically solvable yield *all negative* second order corrections. In fact, both the harmonic oscillator and H-atom perturbations [see (5) and (7)] are such that the *n*-dependence of energy would only be larger. A pure oscillator shows a linear dependence with *n*. But, corrections would be such as to increase the power of *n*. In case of the PB, however, the maximum dependence already exists ($n = 2$). So, either this will prevail, or it can only decrease. Hence, the corrections cannot show a positive n-dependence. The large-*n* behavior of the corrections, if they depend on positive powers of *n*, therefore, can only be negative to satisfy (3).

One can now also explain qualitatively why polarizability of H atom in any state [11 – 13] would be positive, if we remember that the quantity is proportional to volume, i.e., $r^3$, and $r$ goes as $n^2$. A similar conclusion follows for susceptibility that varies as the area, i.e., $r^2$. In both these cases [14], second order energy corrections are therefore likely to be negative for any state.

## 4. Conclusion

In fine, we have justified that there are two extreme classes of perturbation problems. The statistically more probable first class refers to situations with all $E_{k2} < 0$ and inequality (3) is obeyed by them in the $K \to \infty$ limit. For the other class, we have all $E_{k2} > 0$, except for the ground state, and equality (4) is satisfied. The last result is also true of cases where the first few $E_{k2}$ are negative, but subsequent ones all positive. While the first class of problems possesses an immediate practical appeal in respect of response properties [15], the second class (with a wider subclass of problems) is theoretically more fascinating because of the satisfaction of (4), which is typical of finite-dimensional matrix perturbations.